# Combined Extended Rejoinder to "Extended Comment on "One-Range Addition Theorems for Coulomb Interaction Potential and Its Derivatives" by I. I. Guseinov (Chem. Phys., Vol. 309 (2005), pp. 209-213)"

I. I. Guseinov

*Department of Physics, Faculty of Arts and Sciences, Onsekiz Mart University, Çanakkale, Turkey*

*E-mail: isguseinov@yahoo.com*

Submitted to the Los Alamos Preprint Server: 22 August 2007

## Abstract

This article is a thorough critique to the Weniger's comments made to our papers published in prestigious journals in the recent years. A detailed and critical examination of the arguments that led to the suggested comment by Weniger reveals some serious flaws. In our published papers we have shown that the unsymmetrical and symmetrical one-range addition theorems for Slater type orbitals, Coulomb-Yukawa like correlated interaction potentials (CIPs) and their derivatives are derived from the expansions in terms of $\Psi^\alpha$-ETOs that are complete and orthonormal sets of exponential type orbitals in corresponding Hilbert spaces, where $\alpha = 1, 0, -1, -2,...$It should be noted that Lambda and Coulomb Sturmian functions introduced by Hylleraas, Shull and Löwdin which are widely used by Weniger and his coworkers, as indicated by Weniger himself, are the special cases of $\Psi^\alpha$-ETOs for $\alpha = 0$ and $\alpha = 1$, respectively. Thus, the completeness of function sets $\Psi^\alpha$-ETOs in Hilbert spaces suffice to guarantee the existence and convergence of formal series expansions in terms of these functions and, therefore, from a mathematical point of view our treatment of one-range addition theorems is fundamentally flawless. The concrete criticism raised in Weniger's comment against our papers actually touches a very minor aspect of the works that are not relevant at all for the conclusions, which are made. As can be seen from our papers, all of the formulas for different kinds of multicenter integrals over Slater type orbitals with integer and noninteger principal quantum numbers obtained by the use of unsymmetrical and symmetrical one-range addition theorems were tested by computer calculations. We reject the Weniger's personal views about papers published by Guseinov and his coworkers from 1978 to 2006 and respectable referees on one-range addition theorems and multicenter integrals. All claims of inconsistencies and flaws in the theoretical framework are rejected as unfounded. This rejoinder paper contains all of the answers to Weniger's comments.

**Keywords**: Complete orthonormal sets of exponential type orbitals, Hilbert spaces, One-Range addition theorems, Correlated interaction potentials, Multicenter integrals, Noninteger principal quantum numbers

# Contents

# I. Introduction

In the context of atomic and molecular electronic structure calculations, two fundamentally different types of addition theorems occur in the literature. The addition theorems of the first type all have the typical two-range form of the Laplace expansion of the Coulomb potential, which can lead to nontrivial technical problems. These addition theorems can be derived as shown in [1] via rearrangement of 3-dimensional Taylor expansions. There is a second class of addition theorems which can be constructed by expanding a function located at a center $a$ in terms of a complete orthonormal set located at a center $b$. In such an addition theorem, the dependence of the centers $a$ and $b$ is completely contained in the overlap integrals. If the complete orthonormal set occurring in the one-range addition theorem consists of exponentially decaying functions, then its elements can usually be expanded by finite linear combinations of STOs. Consequently, it looks like an obvious idea to express the overlap integrals obtained from the complete orthonormal set in terms of overlap integrals of STOs.

The use of one-range addition theorems in molecular calculations would be highly desirable since they are capable of producing much better approximations than two-range addition theorems. The one-range addition theorems simplify subsequent integrations in multicenter integrals substantially. The one-range addition theorems established in Guseinov's published papers using complete orthonormal sets of $\psi^{\alpha} - ETOs$ [2] could be utilized for the calculation of arbitrary multicenter multielectron integrals, especially for the evaluation of one- and two-electron integrals of HFR equations and multicenter electronic attraction, electric field and electric field gradient integrals occurring in the study of interaction between electrons and nuclei of a molecule.

In the Comment [3,4], Weniger claims that *"Guseinov had failed to understand the mathematical theory behind one-range addition theorems. Thus is bad enough. However, the referees of Guseinov's numerous articles on one-range addition theorems apparently also failed to understand this theory"*. This statement is completely unacceptable. The respectable referees very well understand and examined the published by Guseinov and his coworkers in the years from 1978 to 2006 papers on one-range addition theorems.

The essential facts of Hilbert space and approximation theory as well as all questions of convergence and existence have been taken into account by Guseinov and his coworkers in the context of one-range addition theorems and multicenter integrals. Thus, the Guseinov's treatment of one-range addition theorems is not questionable, and is fundamentally flawless from a mathematical point of view.

The convergence of a large number of one-range addition theorems for Coulomb potential [5], which is the special case of Coulomb-Yukawa like CIPS, was tested for $\alpha = 0$, $\alpha = 1$ and $\alpha = -1$. The accuracy of computer results obtained from one-range addition theorems for Coulomb potential with the different values of $\alpha$ is satisfactory. Guseinov and his coworkers performed the similar tests in all articles. We completely reject the Weniger's personal statements about Guseinov's approach for one-range addition theorems and multicenter integrals over STOs with integer and noninteger principal quantum numbers. It is quite obvious that Weniger had failed to understand the quantum theory behind integer principal quantum numbers. He often contradicts his own previous works.

In this reply, we present the series expansion formulas and computer results for the multicenter integrals of HFR equations for molecules obtained with the help of one-range addition theorems of Coulomb potential. It is demonstrated that the Guseinov's one-range addition theorems for Coulomb potential from a mathematical of view are completely flawless and substantially simplify integration in multicenter integrals.

## 2. Complete Orthonormal Sets of $\Psi^\alpha$-ETOs

It is well known that the exponential type orbitals (ETOs) would be desirable for basis sets in molecular calculations because they can satisfy the cusp condition at the nuclei [6] and the exponential decay for large distances [7-8]. However, the difficulties in calculation of multicenter molecular integrals have restricted the use of ETOs in quantum chemistry. In the literature there is renewed interest in developing efficient methods for the calculation of multicenter molecular integrals by employing ETOs as basis sets [9-17]. Older works mainly using STOs are reviewed in Refs. [18-22].

One of the most promising methods for the evaluation of multicenter molecular integrals is the expansion of STOs in terms of complete orthonormal sets of ETOs placed at shifted center. In Ref. [17] we derived the two kinds of expansion formulas for one-range addition theorems of STOs using so-called Coulomb Sturmian and Lambda ETOs, which into atomic and molecular calculations were introduced in Refs. [23-26], respectively (see also Refs. [27-28]). It should be noted that Coulomb Sturmian and Lambda ETOs are based upon the generalized Laguerre polynomials $L_{n+1}^{2l+1}$ and $L_{n+l+1}^{2l+2}$, respectively. Utilizing relations for these ETOs presented in Refs. [23-26] we are able to find in Ref. [2] the following single analytical formula for the complete orthonormal sets of $\psi^{\alpha}-ETOs$ which are composed of an exponential, a regular solid spherical harmonic, and the non-standard generalized Laguerre polynomials $L_{n+1}^{2l+1}$, $L_{n+l+1}^{2l+2}$, $L_{n+l+2}^{2l+3}$, ...

$$\psi_{nlm}^{\alpha}(\zeta, \vec{r}) = R_{nl}^{\alpha}(\zeta, r)S_{lm}(\theta, \varphi) \tag{2.1}$$

$$R_{nl}^{\alpha}(\zeta, \vec{r}) = (-1)^{\alpha} \left[ \frac{(2\zeta)^3 (q-p)!}{(2n)^{\alpha}(q!)^3} \right]^{1/2} x^l e^{-x/2} L_q^p(x), \tag{2.2}$$

where $x = 2\zeta r$, $p = 2l + 2 - \alpha$, $q = n + l + 1 - \alpha$ and $\alpha = 1, 0, -1, -2, -3, \ldots$. The following relation determines the generalized Laguerre polynomials

$$L_q^p(x) = \sum_{i=1}^{q-p} \beta_{qi}^p x^i, \tag{2.3}$$

where

$$\beta_{qi}^p = (-1)^{p+i}(q-i)! F_i(q) F_{p+i}(q). \tag{2.4}$$

$$F_i(q) = \begin{cases} q! / [i!(q-i)!] & for & 0 \le i \le q \\ 0 & for & i < 0 \quad and \quad i > q \end{cases}. \tag{2.5}$$

The generalized Laguerre polynomials satisfy the orthonormality relation

$$\int_0^{\infty} e^{-x} x^p L_q^p(x) L_{q'}^p(x) dx = \frac{(q!)^3}{(q-p)!} \delta_{qq'}. \tag{2.6}$$

The functions $S_{lm}$ occurring in Eq. (2.1) are the normalized complex ($S_{lm} \equiv Y_{lm}$) or real spherical harmonics. We notice that our definition of phases for the complex spherical harmonics differs from the Condon–Shortley phases [29] by the sign factor. We use phases according to the relation $Y_{lm}^{*}(\theta, \varphi) = Y_{l-m}(\theta, \varphi)$. It should be noted that the Lambda and Coulomb Sturmian ETOs introduced in Refs. [23-26] are the special cases of the $\psi^{\alpha}-ETOs$ for $\alpha = 0$ and $\alpha = 1$, respectively; i.e., $\psi_{nlm}^0 \equiv \Lambda_{nlm}$ and $\psi_{nlm}^1 \equiv \psi_{nlm}$ (see Eqs. (1) and (2) of Ref. [17]).

The $\psi^{\alpha}-ETOs$ are orthonormal with respect to the weight function $(n / \zeta r)^{\alpha}$:

$$\int \psi_{nlm}^{\alpha^{*}}(\zeta, \vec{r}) \bar{\psi}_{n'l'm'}^{\alpha}(\zeta, \vec{r}) dV = \delta_{nn'} \delta_{ll'} \delta_{mm'}, \tag{2.7}$$

where

$$\bar{\psi}_{nlm}^{\alpha}(\zeta, \vec{r}) = \bar{R}_{nl}^{\alpha}(\zeta, r) S_{lm}(\theta, \varphi) \tag{2.8}$$

$$\bar{R}_{nl}^{\alpha}(\zeta, \vec{r}) = \left( \frac{n}{\zeta r} \right)^{\alpha} R_{nl}^{\alpha}(\zeta, r). \tag{2.9}$$

By the representing the generalized Laguerre polynomial in terms of powers of the $x = 2\zeta r$ according to Eq. (2.3) it is easy to obtain for the transformation of $\psi^\alpha -$ and $\bar{\psi}^\alpha - ETOs$ into STOs the expressions

$$\psi^\alpha_{nlm}(\zeta, \vec{r}) = \sum_{n'=l+1}^{n} \omega^{\alpha l}_{nn'} \chi_{n'lm}(\zeta, \vec{r}) \tag{2.10}$$

$$\bar{\psi}^\alpha_{nlm}(\zeta, \vec{r}) = (2n)^\alpha \sum_{n'=l+1-\alpha}^{n-\alpha} \left[ (2n')! / \left[ 2(n'+\alpha) \right]! \right]^{1/2} \omega^{\alpha l}_{nn'+\alpha} \chi_{n'lm}(\zeta, \vec{r}) , \tag{2.11}$$

where the quantities $\omega^{\alpha l}_{nn'}$ are determined by

$$\omega^{\alpha l}_{nn'} = (-1)^{n'-l-1} \left[ \frac{(n'+l+1)!}{(2n)^\alpha (n'+l+1-\alpha)!} F_{n'+l+1-\alpha}(n+l+1-\alpha) F_{n'-l-1}(n-l-1) F_{n'-l-1}(2n') \right]^{1/2} \tag{2.12}$$

Here, the $\chi_{nlm}(\zeta, \vec{r})$ are the normalized STOs defined as

$$\chi_{nlm}(\zeta, \vec{r}) = (2\zeta)^{n+1/2} \left[ (2n)! \right]^{-1/2} r^{n-1} e^{-\zeta r} S_{lm}(\theta, \varphi) . \tag{2.13}$$

We notice that if factorials of negative number occur in these equations, they should be equated to zero, i.e., $\omega^{\alpha l}_{nn'} = 0$ for $n < n'$.

# 3. One-Range Addition Theorems for $\Psi^\alpha$-ETOs, STOs and Coulomb-Yukawa Like CIPs

The aim of this section is to establish the one-range addition theorems for $\psi^\alpha - ETOs$, STOs and Coulomb-Yukawa like CIPs. These addition theorems can be used for the calculation of multicenter multielectron integrals arising in HFR approximation and correlated interaction potentials approaches.

## 3.1. $\Psi^\alpha$-ETOs

To derive the unsymmetrical one-range addition theorems for $\psi^\alpha - ETOs$ we expand the $\psi^\alpha - ETOs$ in terms of $\psi^\alpha - ETOs$ at a displaced center and use Eq. (2.7) for orthonormality relation. Then, we get the desired result,

$$\psi^\alpha_{nlm}(\zeta, \vec{r}_{a1}) = \sum_{n'=1}^{\infty} \sum_{l'=0}^{n-1} \sum_{m'=-l'}^{l'} S^{\alpha^*}_{nlm,n'l'm'}(\zeta, \zeta'; \vec{R}_{ab}) \psi^\alpha_{n'l'm'}(\zeta', \vec{r}_{b1}) , \tag{3.1}$$

where the overlap integrals of $\psi^\alpha - ETOs$ are determined as

$$S^\alpha_{nlm,n'l'm'}(\zeta, \zeta'; \vec{R}_{ab}) = \int \Psi^{\alpha^*}_{nlm}(\zeta, \vec{r}_{a1}) \bar{\Psi}^\alpha_{n'l'm'}(\zeta', \vec{r}_{b1}) dv_1 . \tag{3.2}$$

Using Eqs. (2.10) and (2.11) in (3.2) we obtain for the expansion coefficients in terms of overlap integrals over STOs the following relation:

$$S_{nlm,n'l'm'}^{\alpha}(\zeta,\zeta';\vec{R}_{ab}) = (2n')^{\alpha} \sum_{n''=l+1}^{n} \sum_{\mu'=l'+1-\alpha}^{n'-\alpha} \left[ (2\mu')!/ \left[ 2(\mu'+\alpha) \right]! \right]^{1/2} \omega_{nn''}^{\alpha l} \omega_{n'\mu'+\alpha}^{\alpha l'} S_{n''lm,\mu'l'm'}(\zeta,\zeta';\vec{R}_{ab}), \quad (3.3)$$

where the $S_{n''lm,\mu'l'm'}(\zeta,\zeta';\vec{R}_{ab})$ are the overlap integrals of STOs defined by

$$S_{nlm,n'l'm'}(\zeta,\zeta';\vec{R}_{ab}) = \int \chi_{nlm}^{*}(\zeta,\vec{r}_{a1}) \chi_{n'l'm'}(\zeta',\vec{r}_{b1}) dv_{1}. \quad (3.4)$$

See Sec. 5 for the evaluation of overlap integrals $S_{nlm,n'l'm'}(\zeta,\zeta';\vec{R}_{ab})$.

The formulae for symmetrical one-range addition theorems of $\psi^{\alpha} - ETOs$ are presented in Ref. [30]:

$$\Psi_{nlm}^{\alpha}(\zeta,\vec{r}_{a1}) = \left( \frac{2\pi}{\overline{\zeta}} \right)^{3/2} \sum_{n'=1}^{\infty} \sum_{l'=1}^{n'-1} \sum_{m'=-l'}^{l'} \left( \sum_{N=1}^{\infty} \sum_{L=0}^{N-1} \sum_{M=-L}^{L} D_{nlm,n'l'm'}^{\alpha NLM}(\zeta,\zeta';\overline{\zeta}) \Psi_{NLM}^{\alpha*}(\overline{\zeta},\vec{R}_{ab}) \right) \overline{\Psi}_{n'l'm'}^{\alpha}(\zeta',\vec{r}_{b1}), \quad (3.5)$$

where $\overline{\zeta} = (\zeta + \zeta')/2$ and

$$D_{nlm,n'l'm'}^{\alpha NLM}(\zeta,\zeta';\overline{\zeta}) = \frac{1}{2\pi} (2L+1)^{1/2} C^{L|M|}(lm,l'm') A_{mm'}^{M}$$

$$\times (2N)^{\alpha} \sum_{s=l+1}^{n} \sum_{s'=l'+1}^{n'} \sum_{S=L+1}^{N} \omega_{ns}^{\alpha l} \omega_{n's'}^{\alpha l'} \omega_{NS}^{\alpha L} \frac{[(2(S-\alpha))!]^{1/2}}{[(2S)!]^{1/2}} Q_{sl,s'l'}^{S-\alpha L}(\zeta,\zeta',\overline{\zeta}). \quad (3.6)$$

It should be noted that for $\zeta = \zeta'$, the coefficients $D_{nlm,n'l'm'}^{\alpha NLM}(\zeta,\zeta';\overline{\zeta})$ determined by the relations (3.6) do not depend on the parameters $\zeta$, i.e.,

$$D_{nlm,n'l'm'}^{\alpha NLM} = D_{nlm,n'l'm'}^{\alpha NLM}(\zeta,\zeta,\zeta). \quad (3.7)$$

Thus, Eqs. (3.1) and (3.5) determine all the unsymmetrical and symmetrical one-range addition theorems of $\psi^{\alpha} - ETOs$, respectively.

### 3.2. STOs

In previous works [31-32] the unsymmetrical and symmetrical one-range addition theorems for STOs were derived using complete orthonormal sets of $\Psi^{\alpha} - ETOs$.

In order to obtain the unsymmetrical one-range addition theorems for STOs we expand STOs in terms of complete orthonormal sets of $\Psi^{\alpha}$-ETOs at a displaced center:

$$\chi_{nlm}(\zeta,\vec{r}_{a1}) = \sum_{n'=1}^{\infty} \sum_{l'=0}^{n'-1} \sum_{m'=-l'}^{l'} M_{nlm,n'l'm'}^{\alpha*}(\zeta,\zeta';\vec{R}) \Psi_{n'l'm'}^{\alpha}(\zeta',\vec{r}_{b1}), \quad (3.8)$$

where $\vec{R} = \vec{R}_{ab}$ and

$$M_{nlm,n'l'm'}^{\alpha}(\zeta,\zeta';\vec{R}) = \int \chi_{nlm}^{*}(\zeta,\vec{r}_{a1}) \overline{\Psi}_{n'l'm'}^{\alpha}(\zeta',\vec{r}_{b1}) dv_{1}. \quad (3.9)$$

Here, we have taken into account the orthonormality relation (2.7) for $\Psi^\alpha$-ETOs.

In order to express the right-hand side of eq.(3.8) in terms of STOs we use a particular method suggested in previous publucation[31]. Using this method, we write Eq.(3.8) in the following form:

$$\chi_{nlm}\left(\zeta,\vec{r}_{a1}\right)=\lim_{N\to N_{\max}}\sum_{n'=1}^{N}\sum_{l'=0}^{n'-1}\sum_{m'=-l'}^{l'}M_{nlm,n'l'm'}^{\alpha*}\left(\zeta,\zeta';\vec{R}\right)\Psi_{n'l'm'}^{\alpha}\left(\zeta',\vec{r}_{b1}\right) \qquad (3.10)$$

where $N_{\max}<\infty$ .

Now we rearrange the order of summations in (3.10) using Eq. (2.10) and characteristics of the coefficients $\omega_{nn'}^{\alpha l}$. Then, it is easy to prove for $1\le N<\infty$ the following identity:

$$\sum_{n'=1}^{N}\sum_{l'=0}^{n'-1}\sum_{m'=-l'}^{l'}M_{nlm,n'l'm'}^{\alpha*}\left(\zeta,\zeta';\vec{R}\right)\Psi_{n'l'm'}^{\alpha}\left(\zeta',\vec{r}_{b1}\right)$$
$$=\sum_{n'=1}^{N}\sum_{l'=0}^{n'-1}\sum_{m'=-l'}^{l'}\left(\sum_{n''=n'}^{N}\omega_{n''n'}^{\alpha l'}M_{nlm,n''l'm'}^{\alpha*}\left(\zeta,\zeta';\vec{R}\right)\right)\chi_{n'l'm'}\left(\zeta',\vec{r}_{b1}\right). \qquad (3.11)$$

We take into account Eq.(2.10) in Eqs.(3.9) and (3.11). Then, we obtain finally for the unsymmetrical one-range addition theorems of STOs the following relation:

$$\chi_{nlm}\left(\zeta,\vec{r}_{a1}\right)=\lim_{N\to N_{\max}}\sum_{n'=1}^{N}\sum_{l'=0}^{n'-1}\sum_{m'=-l'}^{l'}V_{nlm,n'l'm'}^{\alpha N*}\left(\zeta,\zeta';\vec{R}\right)\chi_{n'l'm'}\left(\zeta',\vec{r}_{b1}\right) . \qquad (3.12)$$

Here, the expansion coefficients $V^{\alpha N}$ for translation of STOs are determined by

$$V_{nlm,n'l'm'}^{\alpha N}\left(\zeta,\zeta';\vec{R}\right)=\sum_{n''=l'+1}^{N}\Omega_{n''n'}^{\alpha l'}(N)S_{nlm,n''-\alpha l'm'}\left(\zeta,\zeta';\vec{R}\right), \qquad (3.13)$$

where

$$\Omega_{n\kappa}^{\alpha l}(N)=\left[\frac{[2(k-\alpha)]!}{(2\kappa)!}\right]^{\frac{1}{2}}\sum_{n'=\max(n,\kappa)}^{N}(2n')^{\alpha}\,\omega_{n'n}^{\alpha l}\,\omega_{n'\kappa}^{\alpha l} . \qquad (3.14)$$

The quantities $S_{nlm,n'l'm'}$ occurring in Eq. (3.13) are the overlap integrals between the normalized STOs defined by Eq. (3.4)

It can be seen from Eq.(3.13) that the expansion coefficients for the unsymmetrical one-range addition theorems of STOs are expressed through the overlap integrals with STOs.

The symmetrical one-range addition theorems of STOs are given in Ref. [33]:

$$\chi_{\mu\nu\sigma}(\zeta,\vec{r}_{a1})=\frac{1}{\eta^{3/2}}\lim_{\substack{N\to N_{\max}\\N'\to N'_{\max}}}\sum_{n=1}^{N}\sum_{l=0}^{n-1}\sum_{m=-l}^{l}\left[\sum_{u=1}^{N+N'-\alpha+1}\sum_{\nu=0}^{u-1}\sum_{s=-\nu}^{\nu}D_{\mu\nu\sigma,nlm}^{\alpha uus}(N,N';t)\chi_{uus}^{*}(\eta,\vec{R}_{ab})\right]\chi_{nlm}(\eta,\vec{r}_{b1}), \qquad (3.15)$$

where $\eta > 0$ and

$$D_{\mu\nu\sigma,nlm}^{\alpha uvs}(N,N';t) = \sum_{n'=l+1}^{N} \Omega_{nn'}^{\alpha l}(N) \sum_{\mu'=\nu+1}^{N'} g_{\mu'\nu\sigma,n'-\alpha lm}^{\alpha uvs} \sum_{\mu''=\nu+1}^{N'} \frac{(\mu+\mu''-\alpha)!}{\{(2\mu)![2(\mu''-\alpha)]!\}^{1/2}}$$

$$\times \Omega_{\mu'\mu''}^{\alpha\nu}(N')(1+t)^{\mu+1/2}\left(1-t\right)^{\mu''-\alpha+1/2}. \tag{3.16}$$

Here, $t = \dfrac{\zeta-\eta}{\zeta+\eta}$ and $g_{\mu'\nu\sigma,n'-\alpha lm}^{\alpha uvs} \equiv 0$ for $u > \mu' + n' - \alpha + 1$.

The unsymmetrical and symmetrical one-range addition theorems can also be established for noninteger n STOs. For this purpose one should take into account Eq. (3.12) and (3.15) in Eq. (4.27) for the one-center expansion formulas.

### 3.3. Coulomb-Yukawa Like CIPs

It is well known that the determination of multielectron properties for atoms and molecules requires the more accurate solutions of Hartree-Fock (HF) equations [34]. In order to obtain better approximate solutions in HF theory, Hylleraas first introduced the two standard variational approaches in a series of papers on heliumlike systems: (1) the Hylleraas (Hy) method [23,35], in which the interelectronic coordinates $r_{ij}$ are explicitly included in the terms of the wave function; (2) the configuration interaction (CI) method [36,37], in which the wave function is determined by the linear combination of determinantal functions arising from different configurations [38]. There are theoretical grounds [38-39] for thinking that both the CI and the Hy methods are general methods capable of yielding variational solutions that converge to the exact solution of the Schrödinger equation with any desired degree of accuracy if a sufficient number of terms are included. We notice that the CI expansions converge much more slowly than the Hy-method expansions. Recent work on the hybrid technique Hy-CI [40], which avoids many of the complicated integrals, converges rather quickly for small systems. A drawback in the Hy-type expansions, however, is the complexity of the calculation of multicenter multielectron integrals. The Hy method first developed by James and Coolidge [41] has been used for determination of the ground state energy of $H_2$ molecule [42, 43] and is still valid for two- and three-electron atomic and molecular systems (see, e.g., Refs.[44, 45] and references quoted therein).

In this section, using the complete orthonormal sets of $\psi^\alpha - ETOs$ a large number of unsymmetrical and symmetrical one-range addition theorems for Coulomb-Yukawa like CIPs are presented. The addition theorems derived are especially useful for the computation of multicenter multielectron integrals of CIPs occurring in the HFR approximation and explicitly correlated methods.

In Ref. [46], we introduced the CIPs method, in which the positive indices $\mu$, screening parameter $\xi$ and the interelectronic coordinates $x_{ij}$, $y_{ij}$ $and$ $z_{ij}$ are explicitly included in the terms of the CIPs. The combined Coulomb (for $\xi = 0$) and Yukawa (for $\xi > 0$) like CIPs are defined as [32, 46]:

$$f_{\mu\nu\sigma}(\xi, \vec{r}) = f_\mu(\xi, r)\left(\frac{4\pi}{2\nu + 1}\right)^{1/2} S_{\nu\sigma}(\theta, \varphi),$$ (3.17)

where $\mu \geq 0$, $\xi \geq o$ and $S_{\nu\sigma}(\theta, \varphi)$ are the complex (for $S_{\nu\sigma} \equiv Y_{\nu\sigma}$) or real spherical harmonics and

$$f_\mu(\xi, r) = f_{\mu 00}(\xi, r) = r^{\mu - 1} e^{-\xi r}.$$ (3.18)

In order to derive the unsymmetrical one-range addition theorems for Coulomb-Yukawa like CIPs, we expand the function (3.17) in terms of complete orthonormal sets of $\psi^\alpha - ETOs$ at a displaced center. Then, using the method set out in section (3.2) we finally obtain:

$$f_{\mu\nu\sigma}(\xi, \vec{r}_{a1}) = \sqrt{4\pi} \lim_{N \to N_{\max}} \sum_{n=1}^{N} \sum_{l=0}^{n-1} \sum_{m=-l}^{l} W_{\mu\nu\sigma, nlm}^{\alpha N^*}\left(\xi, \eta; \vec{R}_{ab}\right) \chi_{nlm}(\eta, \vec{r}_{b1}),$$ (3.19)

where $\eta > 0$ and

$$W_{\mu\nu\sigma, nlm}^{\alpha N}\left(\xi, \eta; \vec{R}_{ab}\right) = \sum_{n'=l+1}^{N} \Omega_{nn'}^{\alpha l}(N) U_{\mu\nu\sigma, n'-\alpha lm}\left(\xi, \eta; \vec{R}_{ab}\right).$$ (3.20)

Here, $\alpha = 1, 0, -1, -2, \ldots$ and $U_{\mu\nu\sigma, n'-\alpha lm}\left(\xi, \eta; \vec{R}_{ab}\right)$ are the overlap integrals between CIPs and STOs:

$$U_{\mu\nu\sigma, n'-\alpha lm}\left(\xi, \eta; \vec{R}_{ab}\right) = \frac{1}{\sqrt{4\pi}} \int f_{\mu\nu\sigma}^*\left(\xi, \vec{r}_{a1}\right) \chi_{n'-\alpha lm}\left(\eta, \vec{r}_{b1}\right) dv_1.$$ (3.21)

The unsymmetrical one-range addition theorems of Coulomb and Yukawa potentials are obtained from Eq. (3.19) for $\mu = \nu = \sigma = 0$, $\xi = 0$ and $\mu = \nu = \sigma = 0$, $\xi > 0$, respectively,

for Coulomb potential

$$\frac{1}{r_{a1}} = \sqrt{4\pi} \lim_{N \to N_{\max}} \sum_{n=1}^{N} \sum_{l=0}^{n-1} \sum_{m=-l}^{l} W_{000, nlm}^{\alpha N^*}\left(0, \eta; \vec{R}_{ab}\right) \chi_{nlm}(\eta, \vec{r}_{b1})$$ (3.22)

$$W_{000, nlm}^{\alpha N}\left(0, \eta; \vec{R}_{ab}\right) = \sum_{n'=l+1}^{N} \Omega_{nn'}^{\alpha l}(N) U_{000, n'-\alpha lm}\left(0, \eta; \vec{R}_{ab}\right)$$ (3.23)

$$U_{000, n'-\alpha lm}\left(0, \eta; \vec{R}_{ab}\right) = \frac{1}{\sqrt{4\pi}} \int \frac{1}{r_{a1}} \chi_{n'-\alpha lm}\left(\eta, \vec{r}_{b1}\right) dv_1,$$ (3.24)

for Yukawa potential

$$\frac{e^{-\xi r_{a1}}}{r_{a1}} = \sqrt{4\pi} \lim_{N \to N_{\max}} \sum_{n=1}^{N} \sum_{l=0}^{n-1} \sum_{m=-l}^{l} W_{000,nlm}^{\alpha N^*}\left(\xi,\eta;\vec{R}_{ab}\right) \chi_{nlm}\left(\eta,\vec{r}_{b1}\right) \tag{3.25}$$

$$W_{000,nlm}^{\alpha N}\left(\xi,\eta;\vec{R}_{ab}\right) = \sum_{n'=l+1}^{N} \Omega_{nn'}^{\alpha l}(N) U_{000,n'-\alpha lm}\left(\xi,\eta;\vec{R}_{ab}\right) \tag{3.26}$$

$$U_{000,n'-\alpha lm}\left(\xi,\eta;\vec{R}_{ab}\right) = \frac{1}{\sqrt{4\pi}} \int \frac{e^{-\xi r_{a1}}}{r_{a1}} \chi_{n'-\alpha lm}\left(\eta,\vec{r}_{b1}\right) dv_1 . \tag{3.27}$$

The analytical relation for two-center basic Coulomb potential function, Eq. (3.24), is given in previous work [47]

$$U_{nlm}(\eta,\vec{r}) = \frac{2^{n+1}(n+l+1)!}{(2l+1)[(2n)!(2\eta)]^{1/2}(\eta r)^{l+1}} (1 - e^{-\eta r} \sum_{\sigma=0}^{n+l} \gamma_{\sigma}^{l}(n)(\eta r)^{\sigma}) S_{lm}(\theta,\varphi) , \tag{3.28}$$

$$\gamma_{\sigma}^{l}(n) = \frac{1}{\sigma!} - \frac{(n-l)!}{(n+l+1)!(\sigma-2l-1)!} . \tag{3.29}$$

Here $U_{nlm}\left(\eta,\vec{r}\right) \equiv U_{000,nlm}\left(0,\eta;\vec{r}\right)$, $\gamma_{\sigma}^{l}(n) = 0$ *for* $\sigma < 0$ *and* $\sigma > n+l$. In Eq.(3.29) terms with negative factorials should be equated to zero.

We notice that the expression for two-center basic Yukawa potential function, Eq. (3.27), could be obtained by the use of formula for two-center overlap integrals of STOs (see Sec. 5). The formulae for the symmetrical one-range addition theorems of Coulomb-Yukawa like CIPs have been derived in Ref. [33]:

$$f_{\mu\nu\sigma}(\xi,\vec{r}_{21}) = \frac{2^{3/2}}{(2\eta)^{\mu+2}} \lim_{\substack{N \to N_{\max} \\ N' \to N_{\max}}} \sum_{n=1}^{N} \sum_{l=0}^{n-1} \sum_{m=-l}^{l} \left[ \sum_{u=1}^{N+N'-\alpha+1} \sum_{\upsilon=0}^{u-1} \sum_{s=-\upsilon}^{\upsilon} (-1)^{\upsilon} B_{\mu\nu\sigma,nlm}^{\alpha u \upsilon s}(N,N';\xi,\eta) \chi_{u\upsilon s}^{*}(\eta,\vec{r}_{2}) \right] \tag{3.30}$$
$$\times \chi_{nlm}(\eta,\vec{r}_{1}),$$

where $\alpha = 1, 0, -1, -2, \ldots$ and

$$B_{\mu\nu\sigma,nlm}^{\alpha u \upsilon s}(N,N';\xi,\eta) = \left(\frac{4\pi}{2\nu+1}\right)^{1/2} \sum_{n'=l+1}^{N} \Omega_{nn'}^{\alpha l}(N) \sum_{\mu'=\nu+1}^{N'} g_{\mu'\nu\sigma,n'-\alpha lm}^{\alpha u \upsilon s} \sum_{\mu''=\nu+1}^{N'} \frac{(\mu+\mu''-\alpha)!}{\{[2(\mu''-\alpha)]!\}^{1/2}}$$
$$\times \Omega_{\mu'\mu''}^{\alpha\nu}(N') \left(\frac{2\eta}{\xi+\eta}\right)^{\mu+\mu''-\alpha+1} . \tag{3.31}$$

In the case of symmetrical addition theorems for Coulomb like CIPs ($\xi = 0$) the coefficient $B^{\alpha u \upsilon s}$ does not depend on the parameter $\eta$, i.e., $B_{\mu\nu\sigma,nlm}^{\alpha u \upsilon s}(N,N';0,\eta) = B_{\mu\nu\sigma,nlm}^{\alpha u \upsilon s}(N,N')$.

We notice that in our published papers, the series expansion formulae were also derived for the derivatives of unsymmetrical and symmetrical one-range addition theorems of $\Psi^{\alpha} - ETOs$, STOs and Coulomb-Yukawa like CIPs. These derivatives can be useful especially in the study of electric field and its gradient created by the electrons within the molecule. Weniger, however, claims that *"Guseinov did not derive completely symmetrical one-range addition theorems" [3, p.16]*. This statement of Weniger is completely unacceptable. He had failed to understand our theory behind unsymmetrical one-range addition theorems.

## 4. Use of One-Range Addition Theorems for Coulomb Potential in Evaluation of Multicenter Integrals of HFR Equations

The series expansion formulae obtained in Section 3 for the unsymmetrical and symmetrical one-range addition theorems of STOs and Coulomb-Yukawa like CIPs can be used in the derivation of relations for the multicenter integrals of an arbitrary t-electron operator that arises in calculations of atoms and molecules with $N$ electrons, where $2 \le t \le N$. As an example of application, we calculate in this Section, the multicenter integrals of integer and noninteger n STOs appearing in the HFR equations.

### 4.1. Multicenter integrals of integer n STOs

The multicenter integrals over integer $n$ STOs examined in this work have the following form:

One-electron multicenter integrals

$$J_{p_1 p_1'}^{ab,c}(\zeta_1, \zeta_1') = \int \chi_{p_1}^*(\zeta_1, \vec{r}_{a1}) \chi_{p_1'}(\zeta_1', \vec{r}_{b1}) \frac{1}{r_{c1}} dv_1, \qquad (4.1)$$

Two-electron multicenter integrals

$$J_{p_1 p_1', p_2 p_2'}^{ab,cd}(\zeta_1, \zeta_1', \zeta_2, \zeta_2') = \int\int \chi_{p_1}^*(\zeta_1, \vec{r}_{a1}) \chi_{p_1'}(\zeta_1', \vec{r}_{b1}) \frac{1}{r_{21}} \chi_{p_2}(\zeta_2, \vec{r}_{c2}) \chi_{p_2'}(\zeta_2', \vec{r}_{d2}) dv_1 dv_2, \qquad (4.2)$$

where $p_1 \equiv n_1 l_1 m_1$, $p_1' \equiv n_1' l_1' m_1'$, $p_2 \equiv n_2 l_2 m_2$ and $p_2' \equiv n_2' l_2' m_2'$. Here the $\chi_p(\zeta, \vec{r}_{g1})$ denotes that the STO is located at a center g, where $g = a, b, c, d$.

In order to evaluate the one- and two-electron multicenter integrals, Eqs. (4.1) and (4.2), we use the relation (3.22) for the unsymmetrical one-range addition theorems of Coulomb potential in the following form:

$$\frac{1}{r_{21}} = 4\pi \lim_{N \to N_{max}} \sum_{n=1}^{N} \sum_{l=0}^{n-1} \sum_{m=-l}^{l} [\sum_{n'=l+1}^{N} (-1)^l \Omega_{nn'}^{\alpha l}(N) U_{n'-\alpha lm}^*(\eta, \vec{r}_2)] \chi_{nlm}(\eta, \vec{r}_1), \qquad (4.3)$$

where $\alpha = 1, 0, -1, -2...$ Then, we obtain:

$$J_{p_1 p_1'}^{ab,c}(\zeta_1, \zeta_1', \eta) = \sqrt{4\pi} \lim_{N \to N_{max}} \sum_{n=1}^{N} \sum_{l=0}^{n-1} \sum_{m=-l}^{l} (-1)^l S_{p_1 p_1' p}^{abb}(\zeta_1, \zeta_1', \eta) \sum_{n'=l+1}^{N} \Omega_{nn'}^{\alpha l}(N) U_{p'}^*(\eta, \vec{R}_{bc}) \qquad (4.4)$$

$$J_{p_1 p_1', p_2 p_2'}^{ab,cd}(\zeta_1, \zeta_1', \zeta_2, \zeta_2', \eta) = \lim_{N \to N_{max}} \sum_{n=1}^{N} \sum_{l=0}^{n-1} \sum_{m=-l}^{l} (-1)^l S_{p_1 p_1' p}^{abc}(\zeta_1, \zeta_1', \eta) \sum_{n'=l+1}^{N} \Omega_{nn'}^{\alpha l}(N) U_{p', p_2 p_2'}^{c,cd}(\eta, \zeta_2, \zeta_2'), \quad (4.5)$$

where $p \equiv nlm$, $p' \equiv n' - \alpha lm$ and

$$U_{p'}(\eta, \vec{R}_{bc}) = \frac{2^{n'-\alpha+1}(n'-\alpha+l+1)!}{(2l+1)\{[2(n'-\alpha)]!(2\eta)\}^{1/2}(\eta R_{bc})^{l+1}}(1 - e^{-\eta R_{bc}} \sum_{\sigma=0}^{n'-\alpha+l} \gamma_\sigma^l(n'-\alpha)(\eta R_{bc})^\sigma) S_{lm}(\theta_{bc}, \varphi_{bc}) \quad (4.6)$$

$$S_{p_1 p_1' p}^{abc}(\zeta_1, \zeta_1', \eta) = \sqrt{4\pi} \int \chi_{p_1}^*(\zeta_1, \vec{r}_{a1}) \chi_{p_1'}(\zeta_1', \vec{r}_{b1}) \chi_p(\eta, \vec{r}_{c1}) dv_1 \quad (4.7)$$

$$U_{p', p_2 p_2'}^{c,cd}(\eta, \zeta_2, \zeta_2') = \sqrt{4\pi} \int U_{p'}^*(\eta, \vec{r}_{c2}) \chi_{p_2}(\zeta_2, \vec{r}_{c2}) \chi_{p_2'}^*(\zeta_2', \vec{r}_{d2}) dv_2 \quad (4.8)$$

The relationship for two- and three-center overlap integrals of three STOs occurring in Eqs. (4.4) and (4.5), respectively, are presented in Refs. [48] and [49].

Taking into account Eq. (4.6) in (4.8) for $b \to c, c \to 2$ and $\vec{R}_{bc} = \vec{r}_{c2}$ we obtain for the two-center functions $U_{p', p_2 p_2'}^{c,cd}$ the following relation:

$$U_{p', p_2 p_2'}^{c,cd}(\eta, \zeta_2, \zeta_2') = \frac{2^{n'-\alpha}(n'-\alpha+l+1)!}{(2l+1)\{[2(n'-\alpha)]!(2\eta)\}^{1/2}} \frac{N_{n_2 n_2'}(p_2, p_2 t_2)}{p_a^{l+1}} \sum_L (2L+1)^{1/2} C^{L|m_2'|}(lm, l_2 m_2) A_{mm_2}^{m_2'}$$

$$\times \sum_{\alpha\beta q} g_{\alpha\beta}^q(L\lambda_2', l_2'\lambda_2') \begin{cases} G_{-(l+1+\alpha-n_2), n_2'-\beta}^q(p_a; p_2, p_2 t_2) - \sum_{\sigma=l+1+\alpha-n_2}^{n'-\alpha+l} \gamma_\sigma^l(n'-\alpha) p_a^\sigma Q_{\sigma-(l+1+\alpha-n_2), n_2'-\beta}^q(p, pt) & for \quad l+1+\alpha-n_2 > 0 \\ Q_{|l+1+\alpha-n_2|, n_2'-\beta}^q(p_2, p_2 t_2) - \sum_{\sigma=0}^{n'-\alpha+l} \gamma_\sigma^l(n'-\alpha) p_a^\sigma Q_{\sigma+|l+1+\alpha-n_2|, n_2'-\beta}^q(p, pt) & for \quad l+1+\alpha-n_2 \le 0 \end{cases}, \quad (4.9)$$

where $\lambda_2' = |m_2'|$, $p_a = \frac{\eta}{2} R_{cd}$, $p_2 = \frac{R_{cd}}{2}(\zeta_2 + \zeta_2')$, $p_2 t_2 = \frac{R_{cd}}{2}(\zeta_2 - \zeta_2')$, $p = p_a + p_2$, $pt = p_a + p_2 t_2$ and

$$N_{n_2 n_2'}(p_2, p_2 t_2) = \frac{(p_2 + p_2 t_2)^{n_2 + 1/2}(p_2 - p_2 t_2)^{n_2' + 1/2}}{[(2n_2)!(2n_2')!]^{1/2}} \quad (4.10)$$

$$Q_{NN'}^q(p, pt) = \int_1^\infty \int_{-1}^1 (\mu\nu)^q (\mu+\nu)^N (\mu-\nu)^{N'} e^{-p\mu - pt\nu} d\mu d\nu \quad (4.11)$$

$$G_{-NN'}^q(p_a; p, pt) = \int_1^\infty \int_{-1}^1 \frac{(\mu\nu)^q(\mu-\nu)^{N'}}{(\mu+\nu)^N}(1 - e^{-p_a(\mu+\nu)} \sum_{\sigma=0}^{N-1} \frac{[p_a(\mu+\nu)]^\sigma}{\sigma!}) e^{-p\mu - pt\nu} d\mu d\nu \quad (4.12)$$

The analytical and recurrence relations for auxiliary functions $Q_{NN'}^q$ and $G_{-NN'}^q$ are presented in previous paper [50].

Carrying through calculations for the one-center functions $U_{p', p_2 p_2'}^{c,cc}$ analogous to those for the two-center case, we obtain the following formula:

$$U_{p',p_2p_2'}^{c,cc}(\eta,\zeta_2,\zeta_2') = \frac{2^{n'-\alpha}(n'-\alpha+l+1)!}{\{(2l+1)[2(n'-\alpha)]!(2\eta)\}^{1/2}k_2^{l+1}}N_{n_2n_2'}(1,t_2)C^{l|m|}(l_2m_2,l_2'm_2')A_{m_2m_2'}^m$$

$$\times \begin{cases} F_{-(l+1-n_2-n_2')}(k_2) - \displaystyle\sum_{\sigma=l+1-n_2-n_2'}^{n'-\alpha+l} \gamma_\sigma^l(n'-\alpha)k_2^\sigma\frac{[\sigma-(l+1-n_2-n_2')]!}{(k_2+1)^{\sigma-(l-n_2-n_2')}} & for \quad l+1-n_2-n_2' > 0 \\ |l+1-n_2-n_2'|! - \displaystyle\sum_{\sigma=0}^{n'-\alpha+l} \gamma_\sigma^l(n'-\alpha)k_2^\sigma\frac{(\sigma+|l+1-n_2-n_2'|)!}{(k_2+1)^{\sigma+|l+1-n_2-n_2'|+1}} & for \quad l+1-n_2-n_2' \le 0 \end{cases},$$

(4.13)

where $k_2 = \dfrac{\eta}{\zeta_2+\zeta_2'}$ and

$$F_{-N}(k) = \int_0^\infty \frac{1}{x^N}(1-e^{-kx}\sum_{\sigma=0}^{N-1}\frac{(kx)^\sigma}{\sigma!})e^{-x}dx$$

$$= \frac{(-1)^{N-1}}{(N-1)!}[\ln(k+1)+\sum_{i=1}^{N-1}\frac{(-k)^i}{i}].$$

(4.14)

For the small values of k, the function $F_{-N}(k)$ can be calculated by the use of series expansion relation

$$F_{-N}(k) = \sum_{\sigma=0}^\infty \frac{(-1)^\sigma k^{N+\sigma}}{(N-1)!(N+\sigma)}.$$

(4.15)

For the derivation of Eq. (6.15) we have taken into account the following formula [49]:

$$\frac{1}{x^N}(1-e^{-x}\sum_{\sigma=0}^{N-1}\frac{x^\sigma}{\sigma!}) = \sum_{\sigma=0}^\infty \frac{(-x)^\sigma}{(N-1)!\sigma!(N+\sigma)}.$$

(4.16)

From Eqs. (4.4) and (4.5), it can be seen that the three-center nuclear attraction and four-center electron-repulsion integrals of HFR equations are expressed through the two- and three-center overlap integrals of three STOs, respectively, for which the analytical formulas have been established in our previous works [48, 49]. The auxiliary functions $Q_{NN'}^q$ and $G_{-NN'}^q$ occurring in Eq. (4.9) for the two-center functions $U_{p',p_2p_2'}^{c,cd}$, therefore, arising in the case of four-center electron-repulsion integrals have been studied in recently published paper [50]. It should be noted that all the one-, two- and three-center multicenter integras appearing in HFR equations can also be calculated from the formulas (4.4) and (4.5). For this purpose we must calculate one- and two-center overlap integrals of three STOs and use Eqs. (4.9), (4.13), (4.14) and (4.15).

The convergence properties of the series expansion relation for three-center nuclear attraction integral $J_{211,211}^{ab,c}(3.8, 4.6, 4.6)$ obtained by the use of unsymmetrical one-range addition theorems of the Coulomb potential for $\alpha = -1$ are shown in tables 4.1, 4.2 and 4.3. These tables list the partial summations in Eq. (4.4) corresponding to progressively increasing upper summations limits denoted by $N$, $L$ and $M$. As can be seen from tables 4.1 and 4.2, the Eq. (4.4) displays the most rapid convergence to the numerical results with twelve digits stable as a function of summation limits $L$ and $M$. We see that the convergence of the series with respect to $L$ and $M$ is rapid; therefore, we can include only a few terms obtained from the summation over indices $l$ and $m$. Table 4.3 shows that the accuracy of computer calculations obtained in the present algorithm is satisfactory for $N = 15$. Greater accuracy is attainable by the use of more terms in the expansion in Eq. (4.4).

Carrying through calculations for the symmetrical case analogous to those for the unsymmetrical one-range addition theorems, we obtain for the multicenter integrals of HFR equations the following symmetrical formulas:

$$J_{p_1 p_1'}^{ab,c}(\zeta_1, \zeta_1', \eta) = \frac{\sqrt{4\pi}}{\eta^2} \lim_{\substack{N \to N_{\max} \\ N' \to N'_{\max}}} \sum_{n=1}^{N} \sum_{l=0}^{n-1} \sum_{m=-l}^{l} \left[ \sum_{u=1}^{N+N'-\alpha+1} \sum_{v=0}^{u-1} \sum_{s=-v}^{v} B_p^{aq}(N,N') \chi_q(\eta, \vec{R}_{bc}) \right] S_{p_1 p_1' p}^{abc}(\zeta_1, \zeta_1', \eta) \quad (4.17)$$

$$J_{p_1 p_1', p_2 p_2'}^{ab,cd}(\zeta_1, \zeta_1', \zeta_2, \zeta_2', \eta) = \lim_{\substack{N \to N_{\max} \\ N' \to N'_{\max}}} \sum_{n=1}^{N} \sum_{l=0}^{n-1} \sum_{m=-l}^{l} \sum_{u=1}^{N+N'-\alpha+1} \sum_{v=0}^{u-1} \sum_{s=-v}^{v} (-1)^v B_p^{aq}(N,N')$$

$$\times S_{p_1 p_1' p}^{abc}(\zeta_1, \zeta_1', \eta) S_{qp_2 p_2'}^{ccd}(\eta, \zeta_2, \zeta_2') \quad (4.18)$$

where $p \equiv nlm, q \equiv uvs$ and

$$B_p^{aq}(N,N') = \sum_{n'=l+1}^{N} \Omega_{nn'}^{\alpha l}(N) \sum_{\mu'=1}^{N'} g_{\mu'00,n'-\alpha lm}^{\alpha uvs} \sum_{\mu''=1}^{N'} \frac{(\mu''-\alpha)!}{\{[2(\mu''-\alpha)]!\}^{1/2}} \Omega_{\mu'\mu''}^{\alpha 0}(N') 2^{\mu''-\alpha+1} \quad (4.19)$$

The quantities $S^{abc}$ and $S^{ccd}$ occurring in Eqs. ( 4.17) and (4.18) are the multicenter overlap integrals of three STOs. We see from Eq. (4.17) that the multicenter one-electron integrals are expressed through the products of STOs and multicenter overlap integrals. The two-electron multicenter integrals, Eq. (4.18), are the function of the products of multicenter overlap integrals.

Thus, for the calculation of multicenter integrals of HFR equations obtained by the use of symmetrical and unsymmetrical one-range addition theorems of Coulomb potential one can use the formulae for overlap integrals, and auxiliary functions and overlap integrals, respectively.

## 4.2. Multicenter integrals of noninteger $n$ STOs

The multicenter integrals over noninteger $n$ STOs arising in HFR equations are defined as

One-electron multicenter integrals

$$J_{p_1^* p_1'}^{ab,c}(\zeta_1, \zeta_1') = \int \chi_{p_1^*}^*(\zeta_1, \vec{r}_{a1}) \chi_{p_1'}(\zeta_1', \vec{r}_{b1}) \frac{1}{r_{c1}} dv_1, \quad (4.20)$$

Two-electron multicenter integrals

$$J^{ab,cd}_{p_1^*p_1^{'*},p_2^*p_2^{'*}}(\zeta_1,\zeta_1',\zeta_2,\zeta_2') = \iint \chi^*_{p_1^*}(\zeta_1,\vec{r}_{a1})\chi_{p_1'^*}(\zeta_1',\vec{r}_{b1})\frac{1}{r_{21}}\chi_{p_2^*}(\zeta_2,\vec{r}_{c2})\chi_{p_2'^*}(\zeta_2',\vec{r}_{d2})dv_1dv_2, \qquad (4.21)$$

where $p_1^* \equiv n_1^*l_1m_1$, $p_1'^* \equiv n_1'^*l_1'm_1'$, $p_2^* \equiv n_2^*l_2m_2$, $p_2'^* \equiv n_2'^*l_2'm_2'$ and

$$\chi_{n^*lm}(\zeta,\vec{r}) = (2\zeta)^{n^*+1/2}[\Gamma(2n^*+1)]^{-1/2}r^{n^*-1}e^{-\zeta r}S_{lm}(\theta,\varphi). \qquad (4.22)$$

Here, $\Gamma(x)$ is the gamma function [51]. The normalized integer $n$ STOs, Eq. (2.13), can be obtained from Eq. (4.22) for $n^* = n$, where n is an integer.

Taking into account Eqs. (3.22) and (4.3) for the one-range addition theorems of Coulomb potential in Eqs. (4.20) and (4.21) we obtain for the one- and two-electron multicenter integrals of noninteger n STOs the following relations:

$$J^{ab,c}_{p_1^*p_1^{'*}}(\zeta_1,\zeta_1',\eta) = \sqrt{4\pi}\lim_{N\to N_{max}}\sum_{n=1}^{N}\sum_{l=0}^{n-1}\sum_{m=-l}^{l}(-1)^l S^{abb}_{p_1^*p_1'^*p}(\zeta_1,\zeta_1',\eta)\sum_{n'=l+1}^{N}\Omega^{\alpha l}_{nn'}(N)U^*_{p'}(\eta,\vec{R}_{bc}) \qquad (4.23)$$

$$J^{ab,cd}_{p_1^*p_1^{'*},p_2^*p_2^{'*}}(\zeta_1,\zeta_1',\zeta_2,\zeta_2',\eta) = \lim_{N\to N_{max}}\sum_{n=1}^{N}\sum_{l=0}^{n-1}\sum_{m=-l}^{l}(-1)^l S^{abc}_{p_1^*p_1'^*p}(\zeta_1,\zeta_1',\eta)\sum_{n'=l+1}^{N}\Omega^{\alpha l}_{nn'}(N)U^{c,cd}_{p',p_2^*p_2'^*}(\eta,\zeta_2,\zeta_2'),$$

(4.24)

where $U_{p'}(\eta,\vec{R}_{bc})$ is determined by Eq. (4.6) and

$$S^{abc}_{p_1^*p_1'^*p}(\zeta_1,\zeta_1',\eta) = \sqrt{4\pi}\int\chi^*_{p_1^*}(\zeta_1,\vec{r}_{a1})\chi_{p_1'^*}(\zeta_1',\vec{r}_{b1})\chi_p(\eta,\vec{r}_{c1})dv_1 \qquad (4.25)$$

$$U^{c,cd}_{p',p_2^*p_2'^*}(\eta,\zeta_2,\zeta_2') = \sqrt{4\pi}\int U^*_{p'}(\eta,\vec{r}_{c2})\chi_{p_2^*}(\zeta_2,\vec{r}_{c2})\chi^*_{p_2'^*}(\zeta_2',\vec{r}_{d2})dv_2 . \qquad (4.26)$$

With the calculation of these integrals we use the one-center expansion formula for noninteger n STOs in terms of integer n STOs established with the aid of complete orthonormal sets of $\psi^\alpha - ETOs$ [52]:

$$\chi_{n^*lm}(\zeta,\vec{r}) = \lim_{N\to N_{max}}\sum_{n=l+1}^{N}V^{\alpha N}_{n^*l,nl}\chi_{nlm}(\zeta,\vec{r}), \qquad (4.27)$$

where $\alpha = 1,0,-1,-2,...$ and

$$V^{\alpha N}_{n^*l,nl} = \sum_{n'=l+1}^{N}\Omega^{\alpha l}_{nn'}(N)\frac{\Gamma(n^*+n'-\alpha+1)}{[\Gamma(2n^*+1)\Gamma(2n'-2\alpha+1)]^{1/2}} \qquad (4.28)$$

We notice that in the case of integer values of $n^*$ the coefficient $V^{\alpha N}_{n^*l,nl}$ are reduced to the Kronecker symbol, i.e

$$V^{\alpha N}_{n^*l,nl} = \delta_{nn^*}\delta_{Nn^*}. \qquad (4.29)$$

Now we take into account the one-center expansion relation (4.27) in Eqs. (4.25) and (4.26). Then, we obtain the following expressions through the integer n integrals:

$$S_{p_1^* p_1^* p}^{abc}(\zeta_1, \zeta_1', \eta) = \lim_{\substack{N_1 \to N_{1_{max}} \\ N_1' \to N_{1_{max}}'}} \sum_{n_1 = l_1 + 1}^{N_1} \sum_{n_1' = l_1' + 1}^{N_1'} V_{n_1^* l_1, n_1 l_1}^{\alpha N_1} V_{n_1^* l_1', n_1' l_1'}^{\alpha N_1'} S_{p_1 p_1' p}^{abc}(\zeta_1, \zeta_1', \eta) \tag{4.30}$$

$$U_{p', p_2^* p_2^*}^{c, cd}(\eta, \zeta_2, \zeta_2') = \lim_{\substack{N_2 \to N_{2_{max}} \\ N_2' \to N_{2_{max}}'}} \sum_{n_2 = l_2 + 1}^{N_2} \sum_{n_2' = l_2' + 1}^{N_2'} V_{n_2^* l_2, n_2 l_2}^{\alpha N_2} V_{n_2^* l_2', n_2' l_2'}^{\alpha N_2'} U_{p', p_2 p_2'}^{c, cd}(\eta, \zeta_2, \zeta_2') \tag{4.31}$$

As can be seen from the formulas of this section obtained by the use of one-range addition theorems of Coulomb potential, the evaluation of one- and two-electron multicenter integrals over integer and noninteger n STOs is reduced to the calculation of integer n integrals $S^{aaa}, S^{abb}, S^{abc}, U^{c,cd}$ and $U^{c,cc}$. In order to calculate the three-center overlap integrals $S^{abc}$ we should use the one-range addition theorems for STOs.

Thus, all of the integer and noninteger multicenter integrals arising in HFR equations are evaluated by the use of one-range addition theorems for the Coulomb potential and Slater orbitals. It should be noted that the one-range addition theorems for the STOs and Coulomb potential are the special classes of one-range addition theorems obtained by the use of $\psi^\alpha - ETOs$ which belong to the corresponding Hilbert spaces.

*Weniger is very sceptical about all addition theorems for Slater type functions with nonintegral principal quantum number [3, p.28]*. This claim can be rejected with the help of calculation of multicenter integrals over noninteger n STOs for different values of indices $\alpha$. The results of calculation for three-center nuclear attraction integrals over noninteger n STOs, Eq. (4.23), for various values of parameters are presented in Table 4.4. As can be seen from this table that the accuracy of computer results for $\alpha = 0$ and $\alpha = -1$ are satisfactory.

## 5. Two-Center Overlap Integrals

Overlap integrals over integer and noninteger n STOs arise not only in the HFR equations for molecules, but are also central to the calculation of arbitrary multicenter multielectron integrals based on the series expansion formulas obtained by the use of unsymmetrical and symmetrical one-range addition theorems for STOs and correlated interaction potentials which necessitate to accurately calculate the overlap integrals especially for the large quantum numbers.

### 5.1. Overlap Integrals of Integer $n$ STOs

The two-center overlap integrals over integer $n$ STOs with respect to lined-up coordinate systems are defined as

$$S_{nl\lambda,n'l'\lambda}(p,t) = \int \chi_{nlm}^{*}(\zeta,\vec{r}_a)\chi_{n'l'm}(\zeta',\vec{r}_b)dV,  \tag{5.1}$$

where $0 \le \lambda \le l, m = \pm\lambda, p = \dfrac{R}{2}(\zeta+\zeta'), t = (\zeta-\zeta')/(\zeta+\zeta')$ and $\vec{R} \equiv \vec{R}_{ab} = \vec{r}_a - \vec{r}_b$.

We calculate the overlap integrals over integer $n$ STOs using the analytical approach containing well-known auxiliary functions $A_k$ and $B_k$ and the recurrence relations for the basic overlap integrals presented in our previous works [53] and [54, 55], respectively. These expressions are especially useful for computation of overlap integrals on the computer for high quantum numbers, internuclear distances and orbital exponents or vice versa.

In this section, the differences and similarities in organization of existing overlap integral programs are discussed, and a new strategy is developed. This method is computationally simple and numerically well behaved. On the basis of formulas obtained in papers [53-55] we constructed a program for computation of the overlap integrals over integer n STOs using Mathematica 5.0 international mathematical software and Turbo Pascal language packages. The numerical results demonstrate that the computational accuracy of the established formulas is not only dependent on the efficiency of formulas, but also strongly dependent on the used program language packages. Excellent agreement with benchmark results and stability of the technique are demonstrated. Since the overlap integrals over integer n STOs are of considerable importance in the evaluation of arbitrary multicenter integrals by the use of one-range addition theorems, for STOs and potential, it is hoped that the present work will prove useful in tackling more complicated molecular integrals appearing in the determination of various properties for molecules when the HFR approximation is employed.

## A. Analytical Relations in Terms of Auxiliary Functions

In Ref. [53], using the auxiliary function method for the overlap integrals have been established the following formula:

$$S_{nl\lambda,n'l'\lambda}(p,t) = N_{nn'}(t)\sum_{\alpha=-\lambda}^{l}{}^{(2)}\sum_{\beta=\lambda}^{l'}{}^{(2)} g_{\alpha\beta}^{0}(l\lambda,l'\lambda)\sum_{q=0}^{\alpha+\beta}F_q(\alpha+\lambda,\beta-\lambda)$$
$$\times \sum_{m=0}^{n+n'-\alpha-\beta}F_m(n-\alpha,n'-\beta)A_{n+n'-\alpha-\beta-m+q}^{n+n'+1}(p)B_{m+q}(pt),  \tag{5.2}$$

where $A_n$ and $B_n$ are the auxiliary functions defined by [56]

$$A_n(p) = \int_1^\infty \mu^n e^{-p\mu}d\mu = \frac{n!e^{-p}}{p^{n+1}}\sum_{s=0}^{n}\frac{p^s}{s!}  \tag{5.3}$$

$$B_n(pt) = \int_{-1}^{1}\nu^n e^{-pt\nu}d\nu = (-1)^{n+1}A_n(-pt) - A_n(pt)  \tag{5.4}$$

and

$$A_n^k(p) = p^k A_n(p) = n! e^{-p} \sum_{s=k-(n+1)}^{k-1} \frac{p^s}{(s-k+n+1)!} \qquad for \ k \geq n+1 \qquad (5.5)$$

In our previous paper [57], the new analytical relations have been suggested for the fast evaluation of auxiliary functions $A_n$ and $B_n$.

The coefficients $N_{nn'}(t)$ and $F_m(N,N')$ occurring in Eq.(5.2) are determined by

$$N_{nn'}(t) = \frac{[(1+t)]^{n+1/2}[(1-t)]^{n'+1/2}}{\sqrt{(2n)!(2n')!}} \qquad (5.6)$$

$$F_m(N,N') = \sum_{\sigma=\frac{1}{2}[(m-n)+|m-n|)}^{\min(m,N')} (-1)^\sigma F_{m-\sigma}(N) F_\sigma(N') , \qquad (5.7)$$

where $F_m(n) = n!/[m!(n-m)!]$ are the binomial coefficients. It should be noted that, Eq. (5.7) for the generalized binomial coefficients with different notation $D_m^{NN'}$ firstly has been presented by N. Rosen in Ref. [58].

The quantities $g_{\alpha\beta}^0(l\lambda, l'\lambda)$ in Eq.(5.2) are the expansion coefficients for a product of two normalized Legendre functions in elliptic coordinates. The relationship for these coefficients in terms of factorials was given in [59]. In Ref.[60], these coefficients were expressed in terms of binomial coefficients:

$$g_{\alpha\beta}^0(l\lambda, l'\lambda) = \left[ \sum_{i=0}^{\lambda} (-1)^i F_i(\lambda) D_{\alpha+2\lambda-2i}^{l\lambda} \right] D_\beta^{l'\lambda} , \qquad (5.8)$$

where

$$D_\beta^{l\lambda} = \frac{(-1)^{(l-\beta)/2}}{2^l} \left[ \frac{2l+1}{2} \frac{F_l(l+\lambda)}{F_\lambda(l)} \right]^{1/2} F_{(l-\beta)/2}(l) F_{\beta-\lambda}(l+\beta) . \qquad (5.9)$$

## B. Use of Recurrence Relations for Basic Overlap Integrals

In Ref. [55], using the expansion formula for product of two spherical harmonics both with the same center [59], the overlap integrals, Eq.(5.1), were expressed through the basic overlap integrals:

$$S_{nl\lambda, n'l'\lambda}(p,t) = \sum_{l''=\lambda}^{l} \frac{[2p(1+t)]^l}{[2p(1-t)]^{l''}} \left[ \frac{(2l+1)(2l'')! F_{2n'}(2n'+2l'') F_{l''+\lambda}(l+\lambda) F_{l''-\lambda}(l-\lambda)}{(2l''+1)(2l)! F_{2n-2l}(2n)} \right]^{1/2}$$

$$\times \sum_L \sqrt{2L+1} C^L(l'\lambda, l''\lambda) S_{n-l00, n'+l''L0}(p,t), \qquad (5.10)$$

where $C^L(l'\lambda, l''\lambda)$ are the Gaunt coefficients. The basic overlap integrals $S_{n,n'l'} \equiv S_{n00,n'l'0}$ appearing in Eq.(5.10) are determined by the following recurrence relationships:

$$S_{n,n'l'}(p,t) = -a_{l'-1}\left\{ \frac{p(1-t)}{\left[(2n'-1)2n'\right]^{1/2}} S_{n,n'-1l'-1}(p,t) + \frac{\left[(2n'+1)(2n'+2)\right]^{1/2}}{4p(1-t)} S_{n,n'+1l'-1}(p,t) \right.$$

$$\left. - \frac{(1-t)}{4p\left[(1+t)\right]^2}\left[\frac{(2n+1)(2n+2)(2n+3)(n+2)}{(2n'-1)n'}\right]^{1/2} S_{n+2,n'-1l'-1}(p,t) \right\} - b_{l'-1}S_{n,n'l'-2}(p,t), \quad (5.11)$$

where $n \geq 1$, $n' \geq l'+1$, $l' \geq 1$, $-1 < t < 1$ and the coefficients $a_l$ and $b_l$ are determined by

$$a_l = \frac{1}{l+1}\left[(2l+1)(2l+3)\right]^{1/2}, \quad b_l = \frac{l}{l+1}\left(\frac{2l+3}{2l-1}\right)^{1/2}. \quad (5.12)$$

The recurrence relations (5.11) allow us to express $S_{n,n'l'}$ in terms of the integrals $S_{n,n'} \equiv S_{n00,n'00}$ for the calculation of which one can use the following recurrence relations:

for $n \geq 0$, $n' \geq 0$ and $t \neq 0$

$$S_{nn'}(p,t) = \frac{1}{t}\left\{ \sqrt{\frac{n}{2(2n-1)}}(1+t)^2\left[S_{n-1n'}(p,t) - \sqrt{\frac{n-1}{2(2n-3)}}S_{n-2n'}(p,t)\right] - \sqrt{\frac{n'}{2(2n'-1)}}(1-t)^2 \right.$$

$$\times \left[ S_{n,n'-1}(p,t) - \sqrt{\frac{n'-1}{2(2n'-3)}}S_{n,n'-2}(p,t)\right] + \eta_{nn'}(p,t)\left[\delta_{n0}e^{-p(1-t)} - \delta_{n'0}e^{-p(1+t)}\right] \right\}, \quad (5.13)$$

for $n' \geq 0$ and $t = 0$

$$S_{0n'}(p,0) = \left[\frac{n}{2(2n'-1)}\right]^{1/2}S_{0n'-1}(p,0) + \left[\frac{2(2n'+1)}{n'+1}\right]^{1/2}\eta_{0n'+1}(p,0)\,e^{-p}, \quad (5.14)$$

for $1 \leq n \leq n'$ and $t = 0$

$$S_{nn'}(p,0) = \left[\frac{n(2n'-1)}{(2n-1)(n'+1)}\right]^{1/2}S_{n-1n'+1}(p,0) - \left[\frac{n(n-1)(2n'+1)}{2(2n-1)(2n-3)(n'+1)}\right]^{1/2}S_{n-2n'+1}(p,0)$$

$$+ \left[\frac{n'}{2(2n'-1)}\right]^{1/2}S_{nn'-1}(p,0). \quad (5.15)$$

Here,

$$\eta_{nn'}(p,t) = \frac{\left[2p(1+t)\right]^{n+\frac{1}{2}}\left[2p(1-t)\right]^{n'+\frac{1}{2}}}{4p^2\left[(2n)!(2n')!\right]^{\frac{1}{2}}}.$$ (5.16)

With the aid of recurrence relations (5.11), (5.13), (5.14) and (5.15) the basic overlap integrals $S_{n,n'l'}(p,t)$ can be expressed in terms of the functions $S_{00}(p,t)$ and $S_{00}(p,0)$ for the calculation of which we can use the following analytical formulas:

$$S_{00}(p,t) = \frac{1}{t}\eta_{00}(p,t)\left\{e^{-p(1-t)} - e^{-p(1+t)}\right\}$$ (5.17)

$$S_{00}(p,0) = e^{-p}.$$ (5.18)

By the use of calculations we can answer to the following Weniger's comment: *"Moreover, Guseinov should know that an observed agreement of different floating point computations up to a certain number of digits does not necessarily prove anything (see for example [61]) " [3, p.23].* On the basis of Eqs.(5.2) and (5.10), obtained in our papers [53-55], we constructed the programs which were performed in the Mathematica 5.0 international mathematical software and Turbo Pascal 7.0 language packages. The computational results of overlap integrals by the use of Turbo Pascal 7.0 language package program have been examined in our published papers [53-55]. The Barnett's data [61] and results of our calculation using Mathematica 5.0 international mathematical software and Turbo Pascal 7.0 language packages for various values of parameters are represented in Table 5.1. Barnett's data are reproduced by using our scheme with Mathematica while we get different results using the same scheme with Turbo Pascal. Thus, in this paper we show that the discrepancies can be arisen in the case of different programming environments. We note that, the difference between the numerical results of Eqs.(5.2) and (5.10) arise only after forty fifth digits. It should be noted that for the comparison of the accuracy of computer results obtained from the formulas of overlap integrals, one should use the same program language packages.

It is well known from the expert of this field that the problems occur in the evaluation of overlap integrals are as follow: small internuclear distances and small orbital exponents, and high internuclear distances and high orbital exponents. The results of calculation in these cases are given in Table 5.2. As is clear from our tests that the recurrence and analytical formulas presented in this study are useful tool for exact evaluation of the overlap integrals with arbitrary values of quantum numbers, internuclear distances and orbital parameters.

Thus, our program calculates the overlap integrals over STOs with arbitrary quantum numbers $(n, l, n', l', \lambda)$ and variables (p,t).

## 5.2. Overlap Integrals of Noninteger $n$ STOs

The overlap integrals over noninteger $n$ STOs are defined as

$$S_{n^*lm, n''l'm}\left(\zeta, \zeta'; \vec{R}\right) = \int \chi^*_{n^*lm}\left(\zeta, \vec{r}_a\right)\chi_{n''l'm}\left(\zeta', \vec{r}_b\right)dV \, . \tag{5.19}$$

With the calculation of these integrals we use Eq. (4.27) in (5.19). Then, we obtain the series expansion relation in terms of overlap integrals with integer $n$ STOs:

$$S_{n^*lm, n''l'm}\left(p, t\right) = \lim_{\substack{N \to N \\ N' \to N'_{\max}}} \sum_{n=l+1}^{N} \sum_{n'=l'+1}^{N'} V^{\alpha N}_{n^*l,nl} V^{\alpha N'}_{n''l',n'l'} S_{nlm, n'l'm}\left(p, t\right), \tag{5.20}$$

where $p = \dfrac{R}{2}(\zeta + \zeta')$, $t = \dfrac{\zeta - \zeta'}{\zeta + \zeta'}$, $S_{n^*lm, n''l'm}\left(p, t\right) \equiv S_{n^*lm, n''l'm}\left(\zeta, \zeta'; R\right)$

and $S_{nlm, n'l'm}\left(p, t\right) \equiv S_{nlm, n'l'm}\left(\zeta, \zeta'; R\right)$.

For the calculation of overlap integrals over integer $n$ STOs, we can use the analytical formulas and sets of recurrence relations presented in sections (5.1A) and (5.1B). This algorithm is especially useful for the computation of overlap integrals for large quantum numbers of integer $n$ STOs appearing in the series expansion formulas for the multicenter molecular integrals obtained by the use of one-range addition theorems. The overlap integrals with noninteger $n$ STOs, Eqs. (5.20), can be calculated by the use of our computer programs for the overlap integrals over integer $n$ STOs .

Thus, we proposed a new technique for the efficient computation of overlap integrals with noninteger $n^*$ STOs, based on the usage of complete orthonormal sets of $\Psi^\alpha$-ETOs. An analysis of the numerical aspects and several numerical tests confirmed that the convergence and the numerical stability of the relevant formulas are guaranteed. Besides having an excellent convergence rate, the proposed method is perfectly general, valid for arbitrary values of quantum numbers, screening constants and internuclear distances. On the basis of formulae (5.20) we constructed a program for the computation of overlap integrals over noninteger n STOs using Turbo Pascal 7.0 language and Mathematica 5.0 international mathematical software. One can determine the accuracy of computer results obtained in this work for the overlap integrals over noninteger n STOs using different sets of $\Psi^\alpha$-ETOs. The examples of computer calculation are shown in Table 5.3. As can be seen from Table 4 that the calculated values of overlap integrals over noninteger $n$ STOs for $\alpha = 1, 0, -1$ show a good rate of convergence with the literature for the arbitrary values of parameters. Greater accuracy is attainable by the use of more terms in the series expansion formula (5.20). The better accuracies can be obtained, if required, by the use of large number of summation terms.

Table 5.4 lists the partial summations corresponding to progressively increasing upper summation limits of Eq. (5.20) for $N = N'$. We see from this table that the Eq. (5.20) displays the most rapid convergence to the numerical results with eleven digits stable and correct by the 17th terms in the infinite summations.

Weniger claims that "*the rate of convergence of the expansion of a given function in terms of Guseinov's functions $\Psi^\alpha$ may depend quite strongly on the choice of $\alpha$* " *[3, p. 10]*. The answer to this comment can be obtained by the use of calculations. The computational results in the case of overlap integrals over noninteger $n$ STO for various values of indices $\alpha$ are given in table 5.5. As seen from this table, the formulas proposed for the three-center nuclear attraction integrals over STOs can be used in the sets if different values of $\alpha$.

# 6. Rebuttal to "Reply to "Extended Rejoinder to "Comment on "One-Range Addition Theorems for Coulomb Interaction Potential and Its Derivatives" by I. I. Guseinov (Chem. Phys., Vol. 309 (2005), pp. 209-213)", arXiv: 0706.0975v2", arXiv: 0707.3361v1"

Weniger has recently published in arXiv.org a paper (see Ref. [4]) of which the content is almost the same as another paper (see Ref. [3]). In this paper, we have demonstrated that the all Weniger's claims in Ref.[3] are mathematically and computationally wrong. Unfortunately, the incorrect quotations were also reported in the sections of the Ref.[4]. It is easy to see from Refs.[3,4] that the Weniger's claims are personal, not scientific. This aspect has prevented Weniger from understanding the reality of our papers. Therefore, I will not spend any time to reply Weniger if he carries on his wrong claims. I think that the respectable scientists will evaluate the Weniger's inconsistent claims about our papers using my Combined Extended Rejoinder. It should be noted that all of the comments by Weniger have been constructed from the wrong use of our starting point Eq. (3.11).

In Refs. [3] and [4], Weniger claims that *"Guseinov's derivation of his one-range addition theorems for the Coulomb potential [5] is very questionable"; "Guseinov disagreed in his Rejoinder [44, p.7] with my conclusion, and claimed instead that validity of his approach follows from Eq.(3.11) of his Rejoinder."[4, p.7]; "I do not question the validity of Guseinov's Eq. (3.11), but I very much disagree with Guseinov's conclusion that his Eq.*

*(3.11) proves the validity of his rearrangements."[4, p.21]; "There is the practical problem that one-range additions theorems for exponentially decaying functions are fairly complicated mathematical object. Accordingly, explicit proofs of their convergence or divergence are very difficult and would most likely require a considerable amount of time and effort."[4, p.7]; "I am skeptical about the feasibility of Guseinov's approach based on the exclusive use of unsymmetrical addition theorems"[3, p.17]; "I am, however, very skeptical about all addition theorems for Slater-type functions* $\chi_{N,L}^{M}(\beta, \vec{r} \pm \vec{r}')$ *with nonintegral principal quantum number* $N \in \mathbb{R} \setminus N$ *"[3, p.29].*

As is seen from these claims, Weniger has not gone any further in his comments beyond suspicions and nothing about mathematically proving incorrectness of our one-range addition theorems published in Chemical Physics and other journals. These claims are completely disastrous for Weniger. We note that the Eqs.(3.10) and (3.11) in Section 3.2 are the starting point of derivation of one-range addition theorems and their derivatives for $\Psi^{\alpha}$-ETOs , STOs, and Coulomb-Yukawa like CIPs which were published in 1978-2006 years. Unfortunately, Weniger could not see these realities. Therefore, he has these inconsistent claims.

Another Weniger's claims about Eq. (3.11) are the following: *"This step is potentially disastrous. A rearrangement of the order of summations of a double series is not always legitimate and can easily lead to a divergent result"* [3, p.18]; *"Accordingly, Guseinov's manipulations, which produced the rearranged addition theorem (3.12) from (3.10), are dangerous"* [3. pp.32, 33].

The answers to these comments were given in our published papers. So, the closed analytical identity (3.11) for $N < \infty$ is mathematically completely legitimate approach that leads to one-range addition theorems of the type of (3.12) given in this paper. By numerical computations, we additionally have shown in this paper that the expressions proposed in our papers can serve even as a good approximation for the one-range addition theorems in practical applications.

It should be noted that the starting point of Weniger's claims is Eq. (6.8) in Ref [3] obtained for $N = \infty$. This equation is not mathematically legimate and, therefore, is wrong. This wrong step of Weniger is disastrous because he did not understand the exact starting point (Eq. (3.11)) of our papers of one-range addition theorems and their derivatives for $\Psi^{\alpha} - ETOs$ , STOs and Coulomb-Yukawa like CIPs.

We want to comment on the sentence (see Ref.[4, p.1 and p.2]) *"To clarify the situation, the most serious mathematical flaws in Guseinov's treatment of one-range addition theorems are discussed in more depth"; "In this Reply to Guseinov's Rejoinder [44], I discuss once and in more depth the most important mathematical flaws of Guseinov's work. In contrast to my earlier and longer Comment [41], I concentrate entirely on fist-order errors."*.

We again reject the declaration of Weniger, who apparently identifies his personal view of mathematical sciences with mathematics itself. From an academic point of view it seems that there is no reason to discuss these claims, since only personal wrong conceptions are reported in Ref.[4]. We show in this paper that these claims are wrong and the proposed expressions are correct. We, additionally, show by numerical computations that the expressions proposed by Guseinov for the one-range addition theorems are good approximations useful in applications.

Weniger in Ref.[4, p.28] about referees indicates the following: *"Unfortunately, I am not so optimistic. But it would be unfair to blame to exclusively Guseinov's referees. Referreng Guseinov's manuscripts is certainly not easy. It is Guseinov's trademark to produces a large number of short and largely overlaping articles on essentially the same topic. This makes it very hard even for a very competent referee not to get lost in Guseinov's flood of publications and to keep track of Guseinov's truly new results. Moreover, as I know from my own experience as a referee, there is always the temptation to be less critical in the case of a short manuscript than in the case of a (very) long manuscript.*

*In my opinion, part of the problem are short articles. While there can be no doubt that short articles are well suited to present new experimental or computational results, they are basically unsuited for predominantly theoretical or mathematical topics"*.

This idea of Weniger is also completely unjust and wrong. There are a number of competent scientists in this field. It is unfair for Weniger to ignore this fact and say that he himself could be the only very good referee. These scientists encouraged me profoundly to study on one-range addition theorems. It should be noted that all of my articles are original and none of them are alike. Weniger is also wrong in this matter.

In pages 20, 25 and 30 of comment Ref.[3] Weniger indicates that *"Multicenter integrals of exponentially decaying functions are notoriously complicated, and a nice explicit expression for a multicenter integral does not necessarily permit its efficient and reliable evaluation."*

It would be better to ask Weniger why he has performed so many cumbersome studies in this subject even though he knows that the evaluations of molecular integrals of exponentialy decaying functions are notoriously complicated. This shows that there is some kind of paradox.

## 7. Summary and Conclusions

By the use of complete orthonormal sets of $\Psi^{\alpha}$-ETOs the unsymmetrical and symmetrical one-range addition theorems were established in our published papers for STOs and Coulomb-Yukawa like CIPs. Using these theorems, the general formulas in terms of three-center overlap integrals were established for the multicenter $t$-electron integrals that arise in the solution of $N$-electron atomic and molecular problem ($2 \leq t \leq N$) when a correlated interaction potentials approximation in HFR theory is employed. With the help of expansion formulas for translation of STOs, the three-center overlap integrals are expressed through the two-center overlap integrals for the calculation of which the efficient computer programs especially useful for large quantum numbers are available in our group. Therefore, by using the computer programs for the overlap integrals one can calculate the multicenter multielectron integrals of STOs and Coulomb-Yukawa like CIPs appearing in the determination of atomic and molecular multielectron properties when the HFR and explicitly correlated approaches are employed.

It is well known that the series of electronic structure and electron-nuclei interaction properties of a molecule are very sensitive to the minor errors in the wave functions and their derivatives with respect to coordinates of the nuclei [45, 62]. These computational problems, which have to be overcome, depend strongly on the basis functions being used. Therefore, the STOs, which are able to describe correctly the asymptotic behavior of exact solutions of atomic and molecular Schrödinger equation both in the vicinity of the nuclei [6] and at large distances away from the nuclei [7, 8], nowadays play a negligible role in ab initio calculations. It should be noted that, if one tries to study, in particular, the electron-nuclei interactions in molecules with the help of Gaussian type orbitals, the slow convergence of which may lead to serious computational problems. The inherent limitations of Gaussian basis functions necessitate to use the one-range addition theorems for interaction potentials and orbitals for the calculation of matrix elements in the MO LCAO theory with STOs. In our published papers, the one-range addition theorems were also derived for derivatives of STOs and Coulomb-Yukawa like CIPs. These theorems can be used in HFR approximation and explicitly correlated methods, and also in the study of electric field and its gradient induced by electron at the nuclei of a molecule.

It should be noted that the one-range addition theorems presented in our papers have been established with the help of complete orthonormal sets of $\Psi^\alpha$-ETOs by utilizing a non-standard generalized Laguerre polynomials defined by

$$L_q^p(x) = (-1)^p q! \mathcal{L}_{q-p}^p(x) = (-1)^\alpha q! \mathcal{L}_{q-p}^p(x), \tag{6.1}$$

where $\mathcal{L}_{q-p}^p(x) = \mathcal{L}_{n-l-1}^{2l+2-\alpha}(x)$ are the standard generalized Laguerre polynomials which are normally used in special function theory [63]. So, the similar formulas for the one-range addition theorems can also be derived using the following relation for complete orthonormal sets of $\Psi^\alpha$-ETOs in standard convention:

$$\Psi_{nlm}^\alpha(\zeta, \vec{r}) = \left[ \frac{(2\zeta)^3 (q-p)!}{(2n)^\alpha q!} \right]^{1/2} x^l e^{-x/2} \mathcal{L}_{q-p}^p(x) \; S_{lm}(\theta, \varphi). \tag{6.2}$$

The choice of reliable basis atomic orbitals is of prime importance in molecular quantum-mechanical calculations since the quality of molecular properties depends on the nature of these orbitals. It is well known that noninteger $n$ STOs provide a simple but more flexible basis for molecular calculations than integer $n$ STOs [64]. The main problem for the use of noninteger $n$ STOs basis in molecular calculations arises in the evaluation of the multicenter integrals. One of the most promising methods for the evaluation of multicenter molecular integrals with noninteger $n$ STOs is the use of one-range addition theorems for noninteger n STOs (see Section 3.2). With the help of complete orthonormal sets of $\Psi^\alpha$-ETOs the general formulas for the expansion of multicenter integrals of noninteger $n$ STOs through the multicenter integrals over integer $n$ STOs have also been established in our published papers (see Section 4.2.). These relations are useful for the computation of multicenter molecular integrals appearing in the determination of various properties of molecules when noninteger $n$ STOs basis is used in the HFR approximation.

We notice that the origin of the one-range addition theorems and their derivatives for STOs and Coulomb-Yukawa like CIPs with integer and noninteger indices, and $\Psi^\alpha$-ETOs is the Eq. (3.11) the main idea of which was suggested in previous papers [31]. Unfortunately, Weniger did not understand this starting point of our articles that leads to his fundamentally flawed comment about our works on unsymmetrical and symmetrical addition theorems and their applications to multicenter integrals. In Refs [65, 66], the formulas are presented for unsymmetrical and symmetrical one-range addition theorems for STOs and Coulomb – Yukava like correlated interaction potentials of integer and noninteger indices in terms of $\chi$-STOs and $\psi^\alpha$-ETOs, respectively.

**Table 4.1.** Convergence of the series expansion relation for three-center nuclear attraction integral $J_{211,211}^{ab,c}(3.8, 4.6, 4.6)$ as a function of summation limit $L$ for $N = 15$, $R_{ac} = 1.1$, $\theta_{ac} = 60^0$, $\varphi_{ac} = 135^0$, $R_{ab} = 0.1$, $\theta_{ab} = 70^0$, $\varphi_{ab} = 54^0$

| $L$ | $\alpha = -1$ |
|----|----|
| 1 | 8.63162010381515E-01 |
| 2 | 8.48610730270593E-01 |
| 3 | 8.47373159860247E-01 |
| 4 | 8.47489271198125E-01 |
| 5 | 8.47488410946185E-01 |
| 6 | 8.47488072030659E-01 |
| 7 | 8.47488102185615E-01 |
| 8 | 8.47488103395572E-01 |
| 9 | 8.47488103228941E-01 |
| 10 | 8.47488103229998E-01 |
| 11 | 8.47488103229987E-01 |
| 12 | 8.47488103229987E-01 |
| 13 | 8.47488103229987E-01 |
| 14 | 8.47488103229987E-01 |

**Table 4.2.** Convergence of the series expansion relation for three-center nuclear attraction integral $J_{211,211}^{ab,c}(3.8, 4.6, 4.6)$ as a function of summation limit $M$ for $N = 15$, $L = 14$, $R_{ac} = 1.1$, $\theta_{ac} = 60^0$, $\varphi_{ac} = 135^0$, $R_{ab} = 0.1$, $\theta_{ab} = 70^0$, $\varphi_{ab} = 54^0$

| $M$ | $\alpha = -1$ |
|----|----|
| 1 | 8.47488103229987E-01 |
| 2 | 8.47488103229987E-01 |
| 3 | 8.47488103229987E-01 |
| 4 | 8.47488103229987E-01 |
| 5 | 8.47488103229987E-01 |
| 6 | 8.47488103229987E-01 |
| 7 | 8.47488103229987E-01 |
| 8 | 8.47488103229987E-01 |
| 9 | 8.47488103229987E-01 |
| 10 | 8.47488103229987E-01 |
| 11 | 8.47488103229987E-01 |
| 12 | 8.47488103229987E-01 |
| 13 | 8.47488103229987E-01 |
| 14 | 8.47488103229987E-01 |

**Table 4.3.** Convergence of the series expansion relation for three-center nuclear attraction integral $J_{211,211}^{ab,c}(3.8, 4.6, 4.6)$ as a function of summation limit $N$ for $R_{ac} = 1.1$, $\theta_{ac} = 60^0$, $\varphi_{ac} = 135^0$, $R_{ab} = 0.1, \theta_{ab} = 70^0$, $\varphi_{ab} = 54^0$

| $N$ | $\alpha = -1$ |
|---|---|
| 8 | 8.47453482E-01 |
| 9 | 8.47467265E-01 |
| 10 | 8.47486649E-01 |
| 11 | 8.47486881E-01 |
| 12 | 8.47487735E-01 |
| 13 | 8.47487914E-01 |
| 14 | 8.47488122E-01 |
| 15 | 8.47488103E-01 |

**Table 4.4.** Comparison of methods of computing three-center nuclear attraction integrals with noninteger n STO for $\eta = \zeta_1'$ and $N = 12$ .

| $n_1$ | $l_1$ | $m_1$ | $\zeta_1$ | $n_1'$ | $l_1'$ | $m_1'$ | $\zeta_1'$ | $R_{ac}$ | $\theta_{ac}$ | $\varphi_{ac}$ | $R_{ab}$ | $\theta_{ab}$ | $\varphi_{ab}$ | Eq.(4.23) | |
|---|---|---|---|---|---|---|---|---|---|---|---|---|---|---|---|
| | | | | | | | | | | | | | | $\alpha = 0$ | $\alpha = -1$ |
| 1.6 | 0 | 0 | 6.5 | 1.7 | 0 | 0 | 2.3 | 0.2 | 60 | 135 | 0.7 | 150 | 180 | 9.93446587E-01 | 9.93445036E-01 |
| 2.3 | 1 | 0 | 7.6 | 2.5 | 0 | 0 | 4.2 | 0.8 | 36 | 54 | 1.5 | 180 | 200 | 1.47075722E-01 | 1.47076339E-01 |
| 2.8 | 1 | 1 | 4.7 | 2.4 | 1 | 0 | 1.4 | 1.3 | 108 | 90 | 1.1 | 108 | 240 | 2.80477271E-02 | 2.80484428E-02 |
| 2.2 | 1 | 1 | 5.6 | 2.4 | 1 | 1 | 3.5 | 2.5 | 30 | 45 | 0.4 | 120 | 160 | 1.88484873E-01 | 1.88486052E-01 |
| 3.2 | 2 | 1 | 6.4 | 2.2 | 1 | 1 | 5.1 | 2.1 | 126 | 108 | 1.4 | 144 | 260 | 2.65988365E-02 | 2.65993638E-02 |

**Table 5.1.** Comparison with results of Barnett [60]

| $n$ | $l$ | $n'$ | $l'$ | $\lambda$ | $p$ | $t$ | Eqs.(5.2) and (5.10) in Turbo Pascal procedure | Eqs.( 5.2) and (5.10) in Mathematica procedure | Ref.[60] in Mathematica procedure |
|---|---|---|---|---|---|---|---|---|---|
| 3 | 2 | 3 | 2 | 1 | 25 | 0.6 | -4.42287766988261E-04 | -4.4228776698826088067954150E-04 | -4.42287 76698 82608 80679E-04 |
| 4 | 2 | 4 | 3 | 1 | 80 | 0.4 | 4.03505950326382E-17 | 4.0350595032638229810896077E-17 | 4.03505 95032 63822 98108E-17 |
| 5 | 4 | 5 | 4 | 4 | 100 | 0.7 | 1.56200599153976E-14 | 1.5620060274578910374522179E-14 | 1.56200 60274 57891 03745E-14 |
| 7 | 3 | 4 | 3 | 2 | 150 | 0.7 | -1.76861050697887E-18 | -1.7686105069226485908088884E-18 | -1.76861 05069 22648 59080E-18 |
| 9 | 5 | 8 | 4 | 3 | 45 | 0.2 | -5.46510243022867E-08 | -5.4651024302270401738249997E-08 | -5.46510 24302 27040 17382E-08 |
| 10 | 7 | 8 | 2 | 1 | 60 | 0.2 | -1.84189026173558E-10 | -1.8418902617319810642243984E-10 | -1.84189 02617 31981 06424E-10 |
| 10 | 9 | 10 | 9 | 9 | 15 | 0.6 | 6.23122318196866E-04 | 6.2312231819112494647561 02E-04 | 6.23122 31819 11249 46475E-04 |
| 13 | 12 | 13 | 12 | 12 | 25 | 0.01 | 1.35310560392189E-04 | 1.3531057870247123818618 68E-04 | 1.35310 57870 24712 38186E-04 |
| 14 | 13 | 14 | 13 | 13 | 15 | 0.4 | 4.53551312156525E-03 | 4.5355128510679091152303 2E-03 | 4.53551 28510 67909 11552E-03 |
| 15 | 14 | 15 | 14 | 14 | 15 | 0 | 3.74722497038009E-02 | 3.7472249703818919543060 84E-02 | 3.74722 49703 81891 95430E-02 |
| 16 | 15 | 16 | 15 | 15 | 35 | 0 | 1.21686562253236E-06 | 1.2168652185901981885690 61E-06 | 1.21686 52185 90198 18856E-06 |
| 17 | 8 | 8 | 7 | 4 | 50 | 0.1 | -1.00640061354258E-06 | -1.0064006411718817234674 00E-06 | -1.00640 06411 71881 72346E-06 |
| 17 | 16 | 17 | 16 | 16 | 25 | -0.5 | 3.06769565185575E-05 | 3.0677032557901936093803 88E-05 | 3.06770 32557 90193 60938E-05 |
| 18 | 12 | 18 | 12 | 12 | 20 | -0.6 | 6.63931813651240E-05 | 6.6393181369665067751321 20E-05 | 6.63931 81369 66506 77513E-05 |
| 21 | 10 | 9 | 8 | 6 | 45 | 0 | 5.38980685350612E-05 | 5.3898068533814377301727 20E-05 | 5.38980 68533 8143 77301 7E-05 |
| 27 | 8 | 9 | 8 | 7 | 35 | -0.2 | -1.73300982799699E-04 | -1.7442380751969590919366 18E-04 | -1.74423 80751 96959 09193E-04 |
| 30 | 10 | 14 | 10 | 8 | 35 | 0 | 1.35074709592800E-02 | 1.3507470959324333887563 35E-02 | 1.35074 70959 32433 38875E-02 |
| 37 | 8 | 12 | 10 | 6 | 10 | -0.6 | 3.98219849004259E-14 | 3.9822800437709157359620 91E-14 | 3.98228 00437 70915 73596E-14 |
| 40 | 4 | 12 | 4 | 3 | 15 | 0.6 | 9.48379265599810E-02 | 9.4837920832255678538441 9E-02 | 9.48379 22083 22556 78538E-02 |
| 43 | 10 | 18 | 8 | 6 | 60 | -0.4 | -1.15907687123104E-04 | -1.1582565326717481466055 45E-04 | -1.158256 53267 1748 14660E-04 |
| 50 | 4 | 50 | 4 | 4 | 25 | 0.7 | 1.84395901037228E-12 | 1.8439587993243634034031 00208E-12 | 1.84395 87993 24363 40310E-12 |

**Table 5.2.** The comparative values of the two-center overlap integrals over STOs in lined-up coordinate systems for small and high values of integral parameters

| $n$ | $l$ | $n'$ | $l'$ | $\lambda$ | $p$ | $t$ | Eqs.(5.2) and (5.10) in Mathematica procedure | Eqs.(5.2) and (5.10) in Turbo Pascal procedure |
|---|---|---|---|---|---|---|---|---|
| 7 | 4 | 7 | 4 | 4 | 0.01 | 0.01 | 0.999247898270316041412006 | 0.999247898270316 |
| 7 | 4 | 7 | 4 | 4 | 0.1 | 0.001 | 0.9997577667797329229393514 | 0.999757766779732 |
| 7 | 4 | 7 | 4 | 4 | 0.01 | 0.001 | 0.99999015239715781346358966 | 0.999990152397158 |
| 7 | 4 | 7 | 4 | 4 | 0.0 | 0.0 | 1.00000000000000000000000000 | 1.00000000000000000 |
| 7 | 4 | 7 | 4 | 4 | 0.001 | 0.1 | 0.927393290379437884684943 | 0.927393290379438 |
| 8 | 7 | 8 | 7 | 7 | 1E-4 | 1E-4 | 0.99999991470588556843130229397 | 0.999999914705885 |
| 8 | 7 | 8 | 7 | 7 | 1E-6 | 1E-6 | 0.999999999991470588235326272549 | 0.999999999991471 |
| 8 | 7 | 8 | 7 | 7 | 1E-6 | -0.5 | 0.0867003276707393893942732464838 | 0.0867003276707394 |
| 10 | 9 | 10 | 9 | 9 | 1E-8 | 0.6 | 9.2233720368547757939453378486E-03 | 9.22337203685478E-03 |
| 10 | 9 | 10 | 9 | 9 | 0.0 | 0.0 | 1.0000000000000000000000000 | 1.00000000000000000 |
| 10 | 9 | 10 | 9 | 9 | 1E-8 | 1E-8 | 0.999999999999999849761904761905 | 0.99999999999999 |
| 10 | 9 | 10 | 9 | 9 | 1E-5 | -0.8 | 2.19369506403590528994511164E-05 | 2.19369506403591 |
| 12 | 10 | 12 | 10 | 10 | 1E-5 | 1E-5 | 0.999999998748172653525286140114 | 0.999999998748173 |
| 12 | 10 | 12 | 10 | 10 | 1E-6 | 1E-6 | 0.99999999998748172653528112453 | 0.999999999987482 |
| 12 | 10 | 12 | 10 | 10 | 1E-6 | 0.1 | 0.881941811798895655010568341338 | 0.881941811798896 |
| 12 | 10 | 12 | 10 | 10 | 1E-4 | -0.6 | 3.777893185901025124800447871E-03 | 3.77789318590103E-03 |
| 7 | 6 | 7 | 6 | 6 | 50 | 0.1 | 1.460223378297466376711404E-14 | 1.46022337769784E-14 |
| 10 | 9 | 16 | 10 | 9 | 60 | 0.1 | - 4.9132686576421288143263755E-13 | - 4.91327027112068E-13 |
| 10 | 9 | 16 | 10 | 9 | 60 | 0.01 | -1.8109678956664726386189893E-13 | -1.81096834940493E-13 |
| 13 | 10 | 13 | 10 | 10 | 35 | 0.1 | 9.76348508560255594773647305E-07 | 9.76348559116148E-07 |
| 7 | 4 | 7 | 4 | 4 | 100 | 0.1 | 2.72292316027798424617358955E-31 | 2.72292315888289E-31 |
| 75 | 30 | 75 | 20 | 18 | 1E-6 | 0.0 | -8.192975496216878820259263E-78 | -8.19297549621688E-78 |

**Table 5.3.** The comparative values of the two-center overlap integrals over noninteger n STOs in lined-up coordinate systems for various values of parameters and $N = N' = 17$

| $n^*$ | $l$ | $n'^*$ | $l'$ | $m$ | $p$ | $p'$ | $t$ | Eq.(5.20) in Turbo Pascal 7.0 $\alpha = 0$ | Eq.(5.20) in Mathematica 5 | | | Ref.[60] |
|---|---|---|---|---|---|---|---|---|---|---|---|---|
| | | | | | | | | | $\alpha = 0$ | $\alpha = 1$ | $\alpha = -1$ | |
| 7.3 | 4 | 7.3 | 4 | 4 | 2 | 1 | 0.5 | 1.017343149959344E-01 | 1.017343148889628E-01 | 1.017343260081906E-01 | 1.0173435184322346E-01 | 1.101734314960E-01 |
| 3.8 | 0 | 5.5 | 0 | 0 | 2.31 | 1.54 | 11/33 | 2.90802046505438E-01 | 2.908020459831434E-01 | 2.908020430948767E-01 | 2.90802093240919E-01 | 2.90802069369E-01 |
| 5.7 | 1 | 3.8 | 1 | 1 | 2.38 | 1.82 | 4/17 | 8.66889506331727E-01 | 8.668895066998231E-01 | 8.668895064966607E-01 | 8.668895066998231E-01 | 8.66889476942E-01 |
| 7.7 | 4 | 6.6 | 4 | 4 | 6 | 7.5 | -0.25 | 2.34831461718284E-01 | 2.3483146019118006E-01 | 2.3483145709511305E-01 | 2.348314694545564E-01 | 2.34831448531E-01 |
| 4.1 | 2 | 3.7 | 2 | 2 | 10.25 | 9 | 5/41 | 2.93541966880792E-02 | 2.935419701322025E-02 | 2.9354189813759587E-02 | 2.93541978255766E-02 | 2.93217486171E-02 |
| 4.6 | 3 | 3.7 | 2 | 2 | 4 | 2.8 | 0.3 | 3.36298814615661E-01 | 3.362988647194424E-01 | 3.3629890916449356E-01 | 3.3629882528623E-01 | |
| 7.2 | 6 | 7.8 | 6 | 6 | 8 | 7.84 | 0.02 | 1.80791756875938E-01 | 1.8079441278114533E-01 | 1.80794413147871E-01 | 1.8079441276532351E-01 | |
| 8.7 | 4 | 8.8 | 5 | 4 | 0.008 | 6/1250 | 0.4 | -4.50210194347972E-04 | 4.5021019516522707E-04 | 4.502101948196665E-04 | 4.5021019298838096E-04 | |
| 13.2 | 7 | 11.5 | 7 | 6 | 0.06 | 0.054 | 0.1 | 9.84040136524412E-01 | 9.840401364384664E-01 | 9.840401364380328E-01 | 9840401364358113E-01 | |
| 15.5 | 10 | 12.8 | 10 | 10 | 0.06 | 0.054 | 0.1 | | 9.96889730182741E-01 | 9.96889728830528E-01 | 9.968897287341084E-01 | |
| 15.5 | 14 | 15.8 | 14 | 14 | 0.06 | 0.054 | 0.1 | | 8.237701856862741E-01 | 8.237129583400589E-01 | 8.237530045411975E-01 | |

**Table 5.4.** Convergence of the series expansion relations for overlap integrals over noninteger n STOs as a function of summation limits for $N = N'$

| $N$ | Eq.(5.20) for $S_{13.276,11.576}\left(0.06,0.1\right)$ |
|---|---|
| 11 | 0.9840682484939571 |
| 12 | 0.9840407293306768 |
| 13 | 0.9840401363432201 |
| 14 | 0.9840401350076602 |
| 15 | 0.9840401363152588 |
| 16 | 0.9840401364339384 |
| 17 | 0.9840401364384664 |

**Table 5.5.** Convergence of the series expansion relation for overlap integrals over noninteger n STOs as a function of $\alpha$ for $N = N' = 17$

| $\alpha$ | Eq.(5.20) for $S_{15.51010,12.81010}\left(0.06,0.1\right)$ |
|---|---|
| 0 | 0.9968786441071692 |
| 1 | 0.9968786427547844 |
| -1 | 0.9968786426594946 |
| -2 | 0.996878649756326 |
| -3 | 0.9968788855664 |
| -4 | 0.996880222965571 |
| -5 | 0.99688484484723 |